# An open-source software package for on-the-fly deskewing and live viewing of volumetric lightsheet microscopy data

Jacob R. Lamb[1], Edward N. Ward[1] and Clemens F. Kaminski[1]

[1]*Department of Chemical Engineering and Biotechnology, University of Cambridge; Cambridge, UK.*

**Abstract:**

Oblique plane microscopy, OPM, is a form of lightsheet microscopy that permits volumetric imaging of biological samples at high temporal and spatial resolution. However, the imaging geometry of OPM, and related variants of light sheet microscopy, distorts the coordinate frame of the presented image sections with respect to real space coordinate frame in which the sample is moved to navigate to regions of interest. This makes live viewing and practical operation of such microscopes difficult. We present an open-source software package that utilises GPU acceleration and multiprocessing to transform the display of OPM imaging data in real time to produce live views that mimic that produced by standard widefield microscopes. Image stacks can be acquired, processed and plotted at rates of several Hz, making live operation of OPMs, and similar microscopes, more user friendly and intuitive.

**Introduction:**

Fluorescence microscopy is a key tool for biological research. Continual development of more advanced microscopes has helped to drive progress in fields such as cell and developmental biology and in neuroscience[1,2]. Specifically, the study of rapid cellular and subcellular dynamics, requires an ability to take volumetric images with subcellular spatial, and high temporal, resolution.

Selective plane illumination microscopy (SPIM), otherwise known as light sheet microscopy, was first detailed in a landmark paper published in 2004 by Huisken et al[3]. It deployed a novel imaging geometry to allow for 3D imaging with temporal and spatial resolution ideal for probing dynamic processes in full organisms with single cell resolution. Compared to its most similar imaging modality, spinning disk confocal microscopy, SPIM not only offers better temporal resolution but also significantly reduces photo-damage and photobleaching in the sample, thus providing more benign imaging conditions over extended durations[4,5,6]. Consequently, SPIM looked set to replace confocal systems as the imaging modality of choice for volumetric imaging. However, SPIM requires two orthogonal objectives, one to deliver a sheet of excitation light into the sample and a second to image the resultant fluorescent signal (Fig. 1b). As this orthogonal imaging geometry is not usually compatible with inverted sample imaging, where illumination and signal detection take place from underneath the sample, it precludes many of the sample mounting techniques most widely used in biological research, for example the use of multi-well plates and stage mounted incubation chambers[7]. Furthermore, geometrical constraints for positioning the two objectives require that the detection objective has a sufficiently long working distance to prevent its physical obstruction by the body of the illumination objective. This limits the numerical aperture (NA) and thus the achievable spatial resolution is often below that of confocal microscopy. Although SPIM has become the imaging method of choice for large volumetric samples (e.g. whole organisms or 3D tissues), this limitation is the reason the technique has not yet replaced confocal imaging for subcellular, high-resolution studies. In summary, the practical application and adoption of the technology as the method of choice for the majority of biological imaging tasks has remained limited[8].

Variations on the standard SPIM geometry have been developed to address the issues of sample mounting and limitations in usable NA. Open-top SPIM (Fig. 1c) uses two orthogonal objectives below the cover glass to allow for the use of standard sample mounting techniques and lab-on-chip type sample containers. However, due to the dual objective geometry, the usable NA remains limited by the requirement for long optical working distances. Furthermore, a long focal length cylindrical lens is required to correct for astigmatism introduced because of imaging at an angle through a glass coverslip[9,10].

Single objective SPIM (Fig. 1d) is a variant, where a mirror is mounted across the sample to reflect the excitation sheet along the focal plane of the detection objective. This removes restrictions placed on the usable detection NA. However, the addition of the mirror adds significant complexity to the sample mounting process and requires on-the-fly drift correction to ensure the light sheet remains in focus[11,12].

Oblique plane microscopy (OPM) overcomes these shortcomings and is set to revolutionise SPIM as an 'all-round' biological imaging technique.

In OPM, pioneered by Dunsby in 2008[13], a single high NA objective is used in epifluorescence mode, and the system is constructed using a standard inverted microscope configuration, a huge advantage for cell biological imaging (Fig. 1e). The excitation light sheet exits the objective at an oblique angle and the fluorescence signal collected through the same objective. A second objective lens is then used to recreate a 3D image of the fluorescence using aberration free remote refocussing[14]. A third objective, at an angle matched to the sheet, is then used to image the slices through the sample, see Fig. 1.

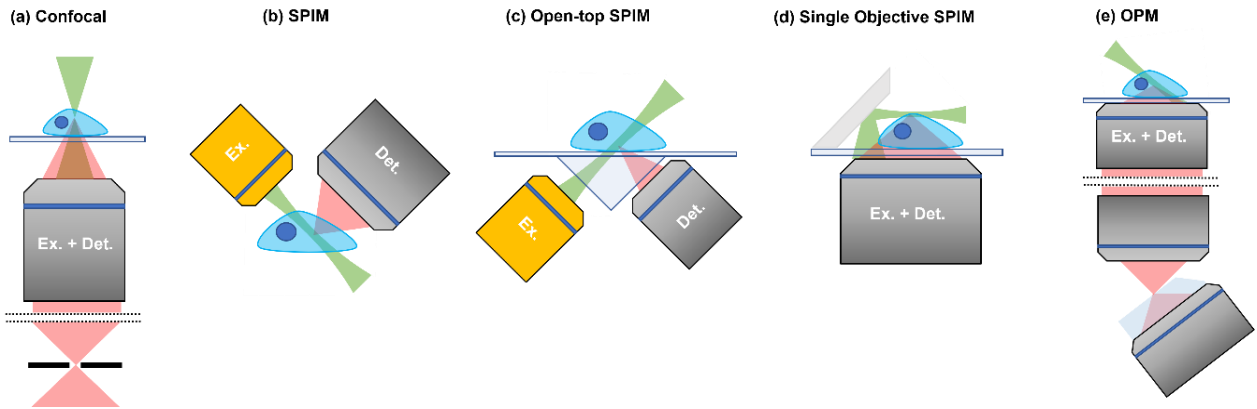

**Fig. 1. Imaging geometries for optical sectioning microscopy:** Optical sectioning is achieved by rejecting fluorescence signal beyond the focal region. In confocal microscopy (a) this is achieved using a pinhole, however, in lightsheet microscopy (b-e) the use of an illumination sheet prevents fluorescence to be generated outside of the focal plane. Conventional two objective based lightsheet modalities (b,c) have restricted spatial resolution due to the requirement for long working distance objectives. Single objective SPIM (d) enables the use of an objective with arbitrary NA, however, the requirement for a mirror in the sample increases mounting complexity. OPM (e) achieves excitation (green) and detection (red) through a single high NA objective placed underneath the sample. Remote refocussing allows the recreation of tilted illumination sheet to be imaged with a tertiary objective at a matching angle.

Since its original publication, OPM has attracted significant interest, and multiple groups are working on further development of the technique. Much of the effort is directed at improving the spatial resolution of OPM. For example, the use of refractive media between the secondary and tertiary objectives permits the full NA of the primary objective to be exploited by refracting the full cone of fluorescence signal captured by the primary objective into the tertiary objective (achieved via the glass block on the tertiary objective as indicated in Fig. 1e)[15,16]. Another improvement concerns the sectioning speed achievable with OPM through use of motorised mirrors[17,18,19]. However, little progress has been made in improving the user friendliness of OPMs, a key requirement for the method to become widely adopted. The angle between imaging and scanning axes produces image sections in OPM that appear distorted and consequently are very difficult to interpret by the operator. Finding regions of interest and optimising imaging conditions becomes very challenging and makes operation of OPM equipment by non-specialists all but impossible.

We present an open-source python-based software package that transforms the display of data obtained by OPM, or indeed any lightsheet microscope where the scan direction is at an angle to the focal plane, into a volumetric projection that mimics a widefield image and is thus easily interpreted by the operator. The software enables the data transformation in real-time and enhances any existing system without laborious modifications to imaging hardware required.

The acquisition of a volume stack with a lightsheet microscope requires sample scanning and camera acquisition to occur in synchronicity. The efficiency with which these operations are synchronised directly impacts the achievable temporal resolution. Our software can operate in two modes. In the first, it works in parallel with the existing software controlling the microscope, and this enables for facile and rapid integration into existing platforms. In the second mode, the software handles all hardware synchronisation directly. For the latter case, optimal speed performance can be achieved.

In this article, we begin by illustrating the geometry of lightsheet imaging and describe the image deskewing process required to recover the true geometry of 3D sample structures from raw lightsheet data. Once deskewed, a maximum projection of the volume presents a mimicked 'widefield' view of the sample. We then show how GPU acceleration and multiprocessing permits the transformation of lightsheet data at rates limited only by the microscope hardware. The software significantly improves the usability of OPM and related imaging modalities, thus making the technique more flexible and easier to use for biological researchers, regardless of user expertise.

**Physical principles for the processing of lightsheet imaging data**

Each lightsheet image represents an angled slice through the sample with a width, length and depth determined by the illumination and detection optics. The angle of the sheet with respect to the scanning axis induces a lateral offset along the length of the image as the sample is scanned (Fig. 2a). This lateral shift, or shear, factor, $l$, between images is given by

$$l = z \cos \alpha, \quad (1)$$

where $\alpha$ is the sheet angle, shown in Fig. 2a, and $z$ is the distance along the scan axis between images. Once a full image stack has been acquired, the data is loaded into image visualisation software, such as Fiji/ImageJ[20]. However, this software is unaware of lateral shifts between each image and so the resultant stack distorts the structure of the sample (Fig. 2b). In order to recover the correct sample geometry, the lateral shifts between sequential images need to be reintroduced before visualisation. This can be done by an affine transformation[18] or, as is done here, the raw images can be remapped into a larger image of dimensions $X$ and $Y$ (Fig. 2c) with the position of each raw image shifting by an amount equal to the original lateral shift between slices. This process is known as deskewing. A maximum intensity projection can then be performed to generate a mimicked 'widefield' view of the data. In the software presented here, by remapping each raw image it is possible to build up the final maximum projection as each image is received from the camera. Conversely, the affine transformation method of deskewing would require a full 3D stack to have been acquired before the data processing could begin. Therefore, the remapping method allows for more efficient parallelisation of the tasks of acquiring and processing the raw images.

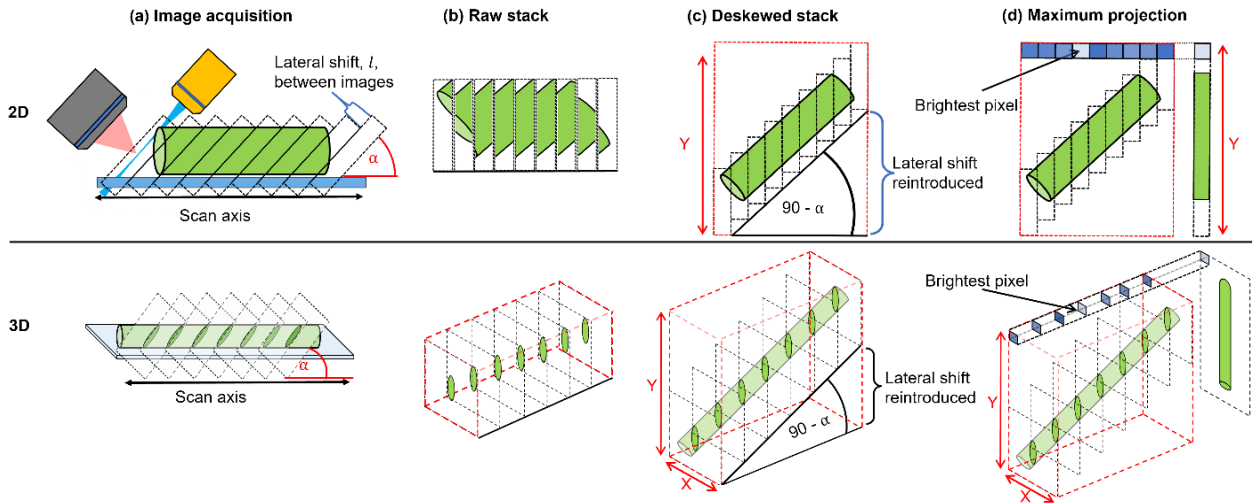

**Fig. 2. Geometries to achieve optical sectioning and corresponding visualisation of data in lightsheet microscopy:** The angle, $\alpha$, between the excitation sheet and the linear displacement along the scan axis causes a lateral offset between each image section. The sample here is represented by a green cylinder. The coverslip is shown in blue. After acquisition the raw stack is loaded into image visualisation software where the lateral offsets are unaccounted for, and the stack thus represents a distorted geometry of the sample. (b) demonstrates how a cylinder would be represented by a raw stack of lightsheet images. The sample has become distorted, and, in 3D, the cylinder cross sections appear to be stretched, forming an oval shape. To regain the correct shape, the lateral shift must be reintroduced in the deskewing process shown in (c). Then, by taking the brightest pixel across the deskewed stack at each pixel position, a maximum projection can be formed, shown in (d).

Due to the detection angle, when the deskewed volume is generated, the sample is reconstructed at an angle to the horizontal equal to $(90 - \alpha)°$, as shown in Fig. 2c. Consequently, the maximum projection of the stack is a projection of the sample when viewed at this angle. Depending on the morphology of the sample, a more intuitive view may require a perspective from a different angle. Rendering such a view requires a rotation of the image coordinate frame.

The shear-warp algorithm is used in the field of computer vision as a fast method of rendering the rotation of 3D volumes. It is based on the principle that, for a given perspective, the projection of a 3D volume when rotated about the centre is equal to the projection of the same volume after it has been sheared and then warped (Fig. 3b)[21]. In the context of deskewed lightsheet data, this means that when the shear factor is set according to Eq. (1), the structure of the sample is faithfully reconstructed. However, if the shear factor is reduced and the maximum projection is warped, it has the effect of rotating the projection angle (Fig. 3a)[22].

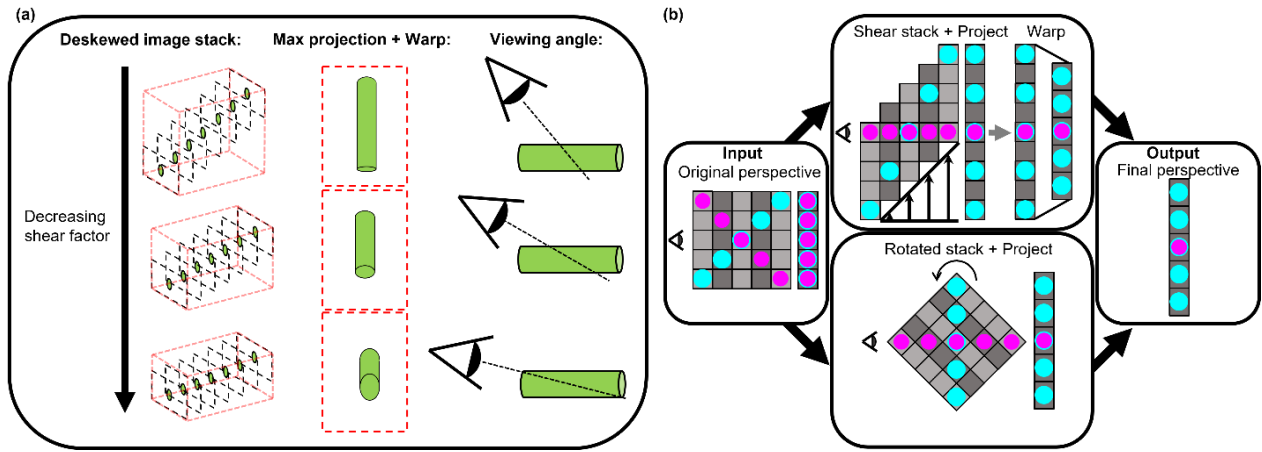

**Fig. 3 The shear warp algorithm enables rapid rotation of lightsheet volumes:** Varying the shear factor ($l$, Eq. 1) when deskewing lightsheet data, then warping the final projected image effectively alters the viewing angle of the sample (a). This is because the projection of a sheared and warped volume is equivalent to the projection of the same volume when rotated around its centre (b).

**Methods for the fast processing and visualisation of lightsheet data**

When imaging using OPMs and other conventional SPIM systems, the live view for users, when searching for the desired imaging location, is a single angled slice through the sample. This view can be very difficult to interpret even for experienced users of lightsheet systems and often the topology of the imaged sample is not apparent until after a full acquisition has been taken and the results have been deskewed. However, the volumetric imaging rate achievable by OPM is on the order of several Hz, which creates a demand for a method to display deskewed image information to the user at a frame rate sufficient to permit facile and intuitive sample navigation during live imaging.

Fast sample scanning rates can be achieved through use of a high-speed galvanometer scanning mirror. Here, imaging rates are only limited by the acquisition speed of the camera. The problem is now to perform the deskewing process quickly enough to allow for real-time display of the data. The fastest way to achieve this is optically. Bo-Jui Chang et al. (2021) report on the use of a matched pair of galvanometer mirrors placed in front of the camera that move synchronously with the sample scanning to project a fully deskewed image across the camera chip during a single exposure[22]. Here, the shear factor applied by the galvanometer mirror pair can be dynamically adjusted to vary the viewing angle in accordance with the shear warp algorithm (Fig. 3). Though very fast, this method is hardware intensive and upgrading an existing OPM system is not trivial. To address this issue, we present a method that performs rapid deskewing in software, that works on any OPM system, and that requires no additional, or modifications to existing, hardware. The software is designed such that it can perform deskewing in real time whilst image acquisition takes place, in single or multiple colour channels simultaneously.

The sequence of operations during imaging involves recording of a stack of images at different scan positions, deskewing each image, calculating the overall maximum intensity projection, and finally displaying the resultant image. During imaging, the live deskewing software needs to control the camera. To ensure widespread support, the package is built on a Python-Micromanager interface (Pycromanager), and thus provides support for the majority of modern scientific cameras[23,24].

To reach frame rates of these operations that are limited only by the imaging hardware, the software must perform the data processing and plotting operations in a time intervals shorter than the camera acquisition time. To achieve this goal, we make use of the Python multiprocessing library and split the different computational tasks into separate processes, so that they can run concurrently.

Utilising GPU optimised code enables camera limited performance, for a single colour channel, with image acquisition, deskewing and plotting split between three separate processes. The image plotting is performed by a separate thread within the process running the GUI. Multi-colour imaging is performed in a similar way, however, here each raw image is split into the separate colour channels with each channel then distributed to a separate deskewing process. In this way the software does not incur significant loss of performance, even if multiple colour channels are acquired simultaneously, and the software can run at frame rates that are only constrained by hardware limitations. The user can adjust the deskewing parameters during imaging and thus the viewing angle, giving a 'live' capability to explore the data from different perspectives. This enables users to gain a better understanding of not only the top-down structure of the sample but also of the full 3D structure prior to data acquisition.

When you have a dim sample, you need long exposure times. The data recorded now comes in so slowly that the live view will refresh too slowly for easy operation of the microscope. We deal with this problem using a so-called rolling update. In this mode, the displayed maximum projection image is updated after every camera exposure on a rolling basis (Fig. 4a).

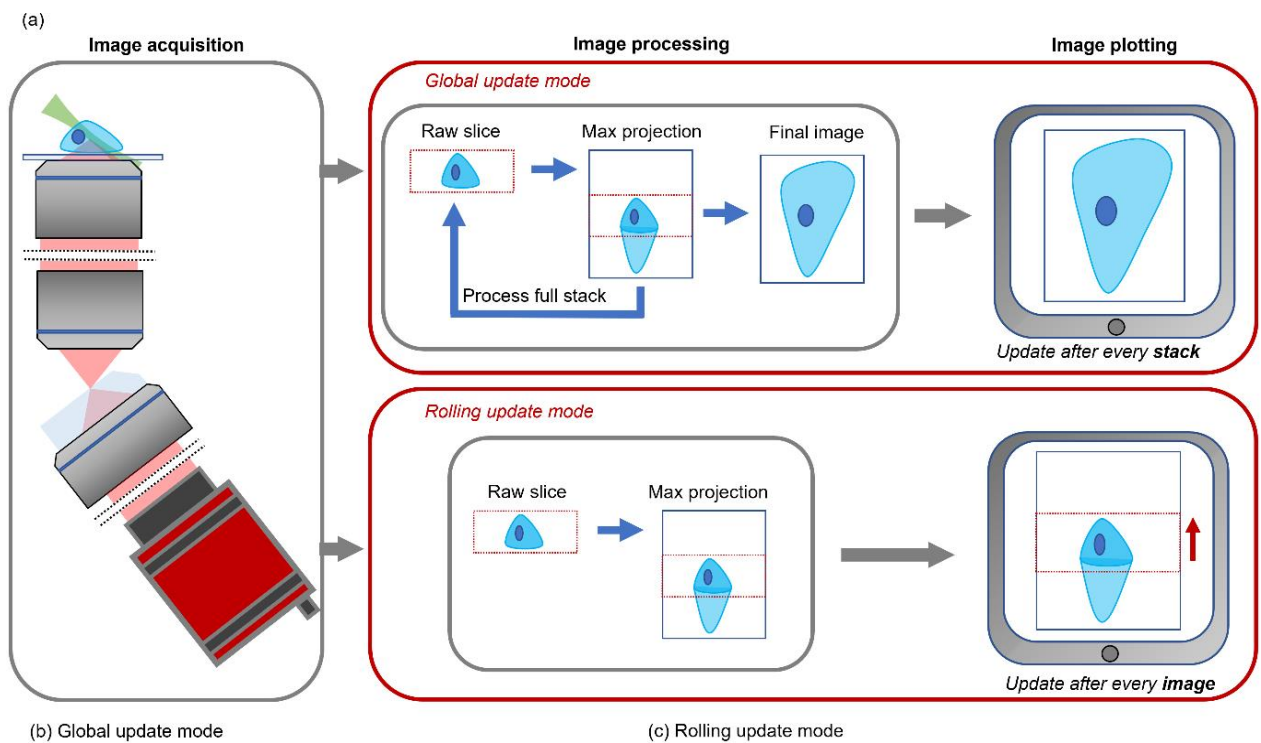

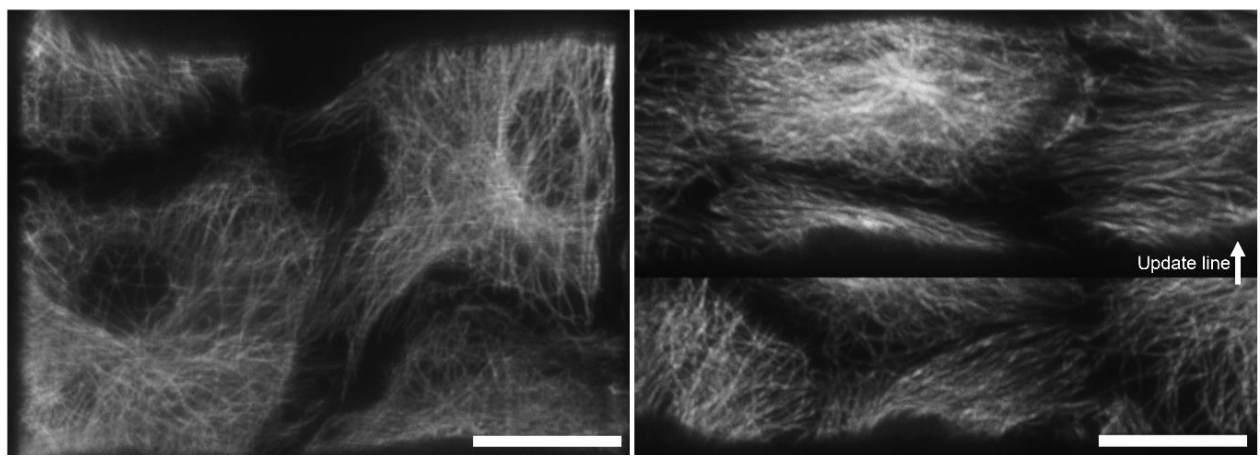

**Fig. 4 Illustration of two different operational modes of the software to permit operation at either low, or fast imaging acquisition speed. For fast frame rates, global updating of the data stack is used. For slow acquisition speeds a rolling update is used.** Camera exposure time is the most influential parameter on software frame rate. (a) Global mode updates the output image after every full stack. However, for long exposure times, this renders the frame rate too low for facile navigation of the sample. Rolling mode updates the output images after each raw image is received from the camera and processed. Global (b) and rolling (c) update modes are shown for the imaging of microtubules in Vero cells. The scale bar is 25um.

The maximum output frame rate of the software is achieved when data processing and plotting are both faster than data acquisition (Supp. Fig. 1). In this scenario, the software is limited only by the microscope hardware. The time taken for the three computational processes (acquisition, processing and plotting) are separately dependent on the parameters of exposure time, number of images per stack and field of view (FOV) along the scan axis, see Supp. Table 1.

For large field of views, the size of the final projected image (see Y in Fig. 2c) increases, and both the image processing (deskewing then projecting) and plotting tasks require additional processing time. However, if one assumes that the number of images per stack remains constant and instead the step size between images increases, the camera acquisition time remains constant. For a camera exposure time of 0.1 ms, a region of interest of 1304 × 87 pixels and 50 images per stack, the output frame update rate remains hardware limited, at 12.5 fps, until the displayed image reaches a size of 3652 × 1304 pixels (420 × 150 μm for a pixel size of 115 nm), at which point image plotting becomes the slower process (Supp. Fig. 2).

**Conclusions:**

We present a data transformation and visualisation tool to enhance the usability of lightsheet microscopes. The software outputs views of projected lightsheet data volumes to facilitates sample navigation and general microscope operation. The software is ready-to-use for integration with new or existing lightsheet microscopes without need for laborious and costly hardware upgrades.

The software utilises GPU acceleration and multiprocessing to process and display live data at rates limited only by the hardware of the microscope. Choosing between a global and a rolling update mode allows the user to optimise the output frame rate depending on camera exposure time.

We demonstrate the software on a state-of-the-art oblique plane microscope, capable of capturing volumetric image stacks containing 100s of slices at repetition rates of several Hz. However, the software is compatible with all lightsheet modalities where the illumination sheet is at an angle to the scan axis. The source code for the software is freely available on GitHub with full instructions on how to implement and use the system on a specific lightsheet system.

**Code availability**

The source code for the live deskewing software presented here is freely available at https://github.com/Jrl-98/Live-Deskewing

**Acknowledgements**

The authors would like to thank Luca Mascheroni for providing fixed cells for imaging and Rebecca McClelland for her helpful comments on the manuscript.

# An open-source software package for on-the-fly deskewing and live viewing of volumetric lightsheet microscopy data

Jacob R. Lamb[1], Edward N. Ward[1] and Clemens F. Kaminski[1]

[1]Department of Chemical Engineering and Biotechnology, University of Cambridge; Cambridge, UK.

## Supplementary Information:

**Supplementary Videos:** Videos available at: https://github.com/Jrl-98/Live-Deskewing/blob/main/README.md

| | |
|---|---|
| **Supplementary Video 1** | Video showing a practical demonstration of deskewing software on an oblique plane microscope (OPM). The software provides live views of microtubules and actin filametns imaged in Vero cells. In this example the software is operated in global update mode, suitable for experiments, where image recording rates are very fast. |
| **Supplementary Video 2** | Video showing a practical demonstration of deskewing software on an oblique plane microscope (OPM). The software provides live views of microtubules and actin filametns imaged in Vero cells. In this example the software is operated in rolling update mode, suitable for experiments, where image recording rates are slow. |

**Supplementary Figures:**

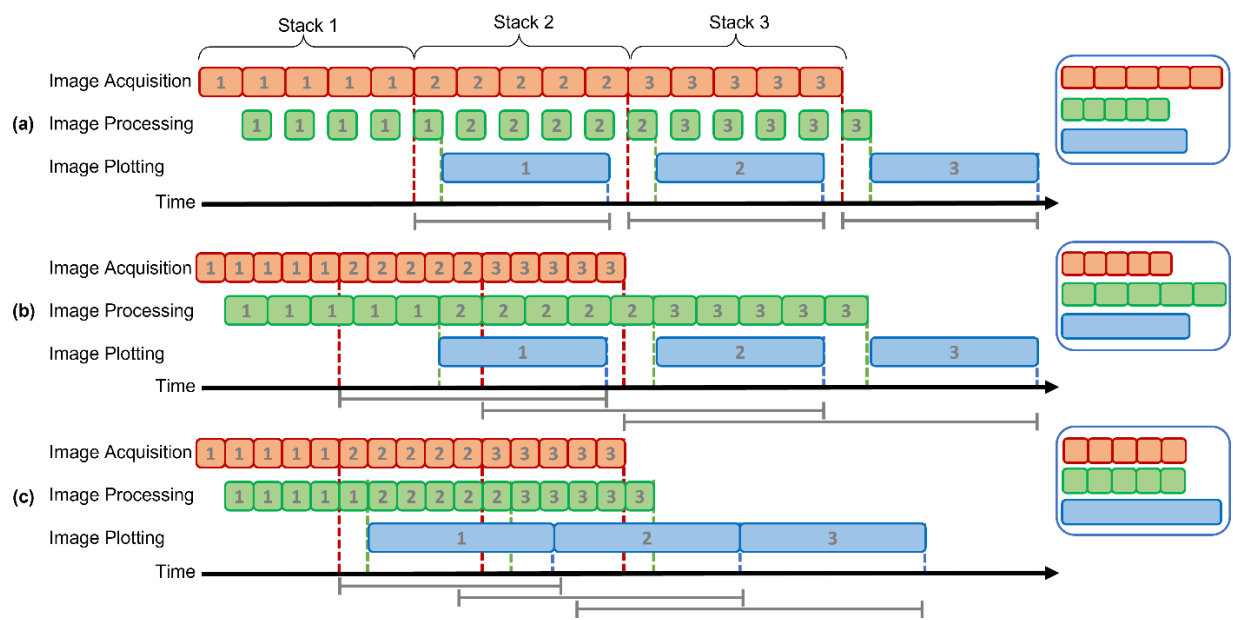

**Supplementary Figure 1: An illustration of how software performance is affected by image acquisition speed.** A stack of raw images must be recorded and processed before the final projection is plotted. The process flow is split into three conncurent processes: image acquistion (orange), image processing (green), and image plotting (blue). (a) When image acqusition is the slowest process, refresh rates for the data display are hardware limited. When either image processing (b) or image plotting (c) are the slowest processes, the camera acquires images faster than the software is able to handle them. Under these conditions, the time between the final image of the stack being acquired and the projection being plotted (grey bar beneath the time axis) grows with each stack. Consequently, the lag time between user navigation and image output increases, resulting in suboptimal user experience.

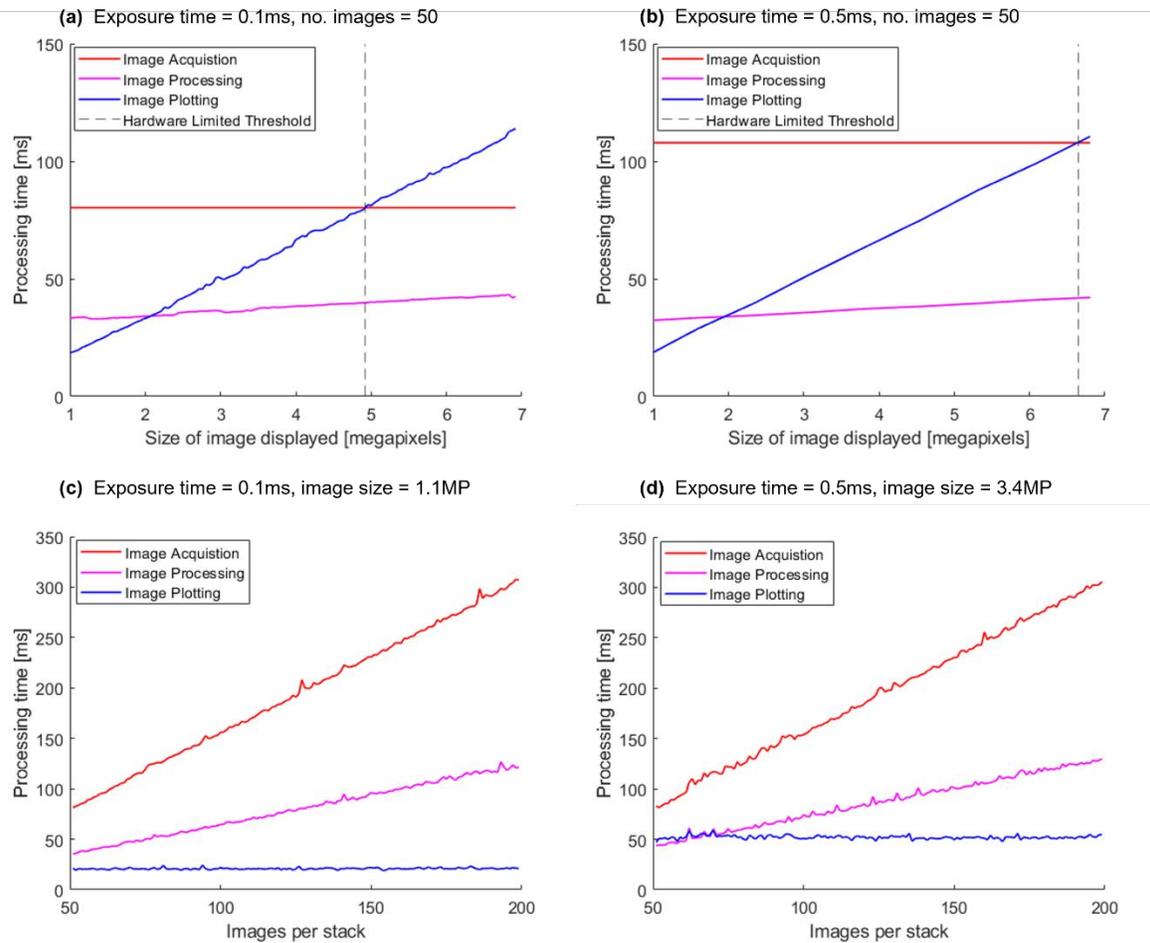

**Supplementary Figure 2: Software performance only remains hardware limited whilst image acquisition is the slowest process. Here the hardware limited threshold is shown for 4 different sets of imaging parameters.** The times required for image acquisition, processing, and plotting each depend differently on the displayed image size, the camera exposure time, and number of images acquired per stack. The software refresh rate remains limited by hardware performance only when the physical image acquisition is the longest task in the process queue. All plots here were produced using a region of interest spanning 1304x87 pixels on the camera. (a,b) For a given exposure time and number of images per stack, image acquisition remains constant, however, image processing and plotting increase linearly. The software stops being hardware limited at an image size of 4.9MP and 6.8MP for exposure times of 0.1ms and 0.5ms, respectively. Increasing the exposure times further would increase the maximum displayed image size where the software remains hardware limited. (c,b) For a given exposure time and image size, image plotting remains constant, however, image acquisition and processing times increase linearly.

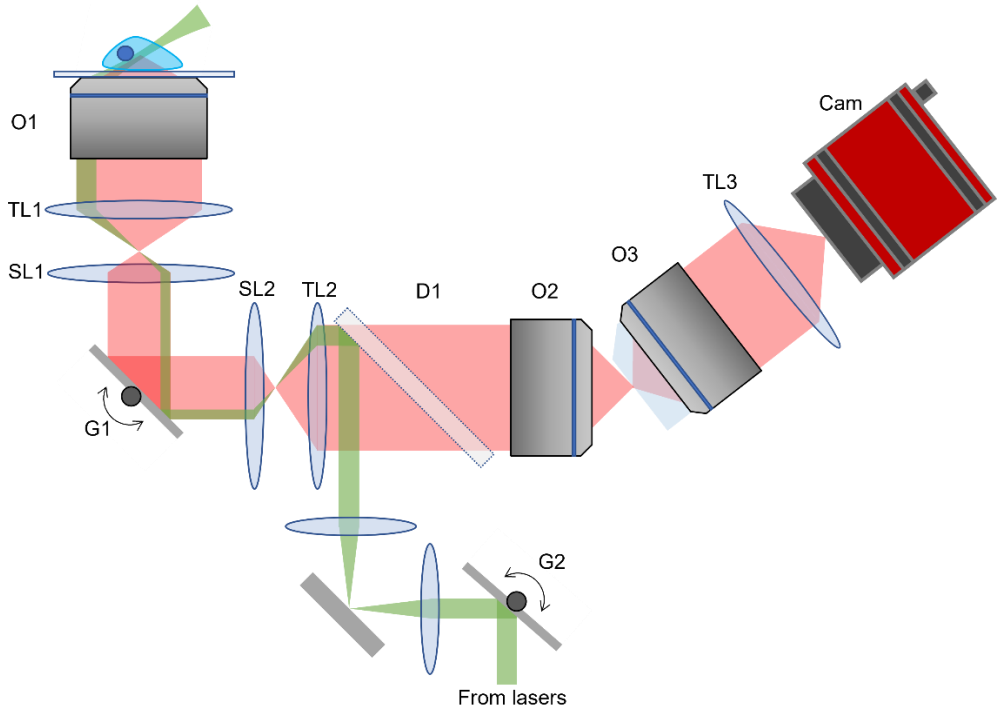

**Supplementary Figure 3: Optical layout for the oblique plane microscope used in this paper:** The primary objective O1 (Olympus UPLSAPO60XS2) collects the fluorescent signal. A tube lens, TL1 (Olympus SWTLU-C), then forms an image of the sample. This image plane is then relayed in a 4f system by two identical scan lenses, SL1 and SL2 (Thorlabs AC-508-075-A and AC-254-100-A combined in a Plossl configuration). In the back focal plane (BFP) at the centre of this 4f system a galvanometric mirror (Scanlab dynAXIS M) is positioned to allow for mirror-based sample scanning. A tube lens with focal length of 193mm, TL2 (combined Thorlabs AC-508-1000-A and TTL200), is then used to perfectly map the BFP of O1 onto the BFP of O2 (UPLXAPO40X objective) to form a remote refocussing system. An angled third objective, O3 (AMS-AGY v1), collects the fluorescent signal and a tube lens, TL3 (Throlabs TTL200), creates an image on the camera (Photometrics Kinetix 22). The sheet is formed by rapidly scanning the focused beam. This is achieved by rotating the beam in a BFP using a galvanometric mirror, G2 (Cambridge Technology 6215H).

## Supplementary Notes:

**Supplementary note 1: Hardware control and synchronisation**

The temporal resolution of galvo scanning OPM is theoretically limited only by the camera acquisition time. However, without finely optimised hardware synchronisation between camera exposures and galvo movements, this theoretical limit cannot be reached, thus losing temporal resolution unnecessarily.

The software has the functionality to control galvo scanning, lasers, filters and external camera triggering, in a standalone package to control the full OPM system. Anyone constructing an OPM can make use of it 'out of the box' without the need to write complex trigger and synchronisation algorithms.

The galvanometer scan position is controlled by an analogue voltage. This voltage can either be written after each camera exposure or preloaded and externally triggered by the camera. Externally tiggering the scan voltages from the exposure line of the camera allows the galvo to move and settle during the camera readout time, therefore not limiting the frame rate. Lasers can be controlled by either an analogue or digital signal. Both the filter wheel position and camera triggering are controlled using digital signals. All the hardware can be set up within an easy-to-use interface and configuration files can be created to allow fast hardware initialisation.

Should the user wish to use their own software to control the hardware of the system, the software can still fully function with any subset of the hardware under its control, provided that the camera is initialised in micromanger. With full access to the source code, users can easily develop support for any other hardware they may require and seamlessly integrate it with the software.

**Supplementary note 2: Cells**

Vero cells (from the American Type Culture Collection, CCL-81) were cultured under standard conditions (37°C and 5% CO2) in Dulbecco's minimum essential medium (DMEM, Sigma-Aldrich) supplemented with 10% heat-inactivated foetal bovine serum (Gibco), antibiotics/antimycotics [penicillin (100 U/ml), streptomycin (100 μg/ml), and amphotericin B (0.025 μg/ml), Gibco], and 2 mM L-glutamine (GlutaMAX, Gibco).

**Supplementary note 3: Immunostaining**

Vero cells were seeded in a glass-bottom 8-well μ-slide (Ibidi, 80827) at a density of 20,000 cells per well and cultured under standard conditions. After 24 hours, cells were fixed by incubation with 4% methanol-free formaldehyde (ThermoFisher Scientific, 28906) and 0.1% glutaraldehyde (Merck, 340855) in cacodylate buffer (100 mM, pH 7.4) for 15 minutes at 37°C. Cells were then washed three times with PBS and permeabilized by incubation with a 0.2% solution of Triton X-100 in PBS for 15 minutes at room temperature. Unspecific binding was blocked by incubating with 10% goat serum (Abcam) and 100 mM glycine in PBS for 30 minutes at room temperature. Without washing, the samples were incubated with an anti-beta-tubulin mouse primary antibody (Abcam, ab131205) diluted 1:200 in PBS containing 2% goat serum for 1 hour at room temperature. After three washes in PBS, the samples were incubated with an Alexa Fluor 568-conjugated anti-mouse IgG1 goat secondary antibody (ThermoFisher Scientific, A-11011) diluted 1:400 in PBS containing 2% goat serum for 1 hour at room temperature in the dark. Samples were then washed 3 times and incubated with a 150 nM solution of ATTO 647N-conjugated phalloidin (Sigma Aldrich, 65906) in PBS for 30 minutes at room temperature in the dark. Cells were then washed three times with PBS and kept in PBS containing 0.05% sodium azide at 4°C until the moment of imaging

**Supplementary note 4: Computer hardware**

All results presented in the manuscript were obtained running the deskewing software on a computer with an AMD Ryzen 9 5900X 12-Core 3.70 Ghz Processor, 48.0 GB of RAM and a Nvidia RTX 3060 GPU.

**Supplementary Table:**

**Supplementary table 1: Each of the software processes are individually dependent on the user control variables**

|  | Image acquisition time | Image processing time | Image plotting time |
| --- | --- | --- | --- |
| **Increasing camera exposure time** | Increasing | Invariant | Invariant |
| **Increasing number of images per stack** | Increasing | Increasing | Invariant |
| **Increasing field of view along the scan axis** | Invariant | Increasing | Increasing |